\documentclass[twocolumn,showpacs,preprintnumbers,amsmath,amssymb]{revtex4}

\usepackage{graphicx}
\usepackage{dcolumn}
\usepackage{bm}

\begin{document}

\preprint{APS/123-QED}

\title{Collective behavior of "electronic fireflies"}
\author{M. Ercsey-Ravasz$^{1,2}$, Zs. S\'ark\"ozi$^1$,  Z. N\'eda$^1$, A. Tunyagi$^1$, and I. Burda$^1$} 
\affiliation{$^1$ Babe\c{s}-Bolyai University, Faculty of Physics, RO-400084, Cluj, Romania \\
     $^2$ P\'eter P\'azm\'any Catholic University, Faculty of Information Technology,   HU-1083, Budapest, Hungary }

\date{\today}

\begin{abstract}
A simple system composed of electronic oscillators capable of emitting and detecting light-pulses is studied.
The oscillators are biologically inspired, their behavior is designed for keeping a
desired light intensity, $W$, in the system.  From another perspective, the system behaves like
modified integrate and fire type neurons that are pulse-coupled with inhibitory type interactions:
the firing of one oscillator delays the firing of all the others. Experimental and computational studies 
reveal that although no driving force favoring synchronization is considered, for a given interval of $W$
phase-locking appears. This weak synchronization is sometimes accompanied by complex
dynamical patterns in the flashing sequence of the oscillators.

\end{abstract}

\maketitle

\section{Introduction}

Synchronization of quasi-identical coupled oscillators is 
one of the oldest and most fascinating problems in physics \cite{strogatz,stewart,strogatz2}. Its history 
goes back to C. Huygens who first noticed the synchronization of pendulum clocks hanging
on the same wall. Besides
mechanical or electric oscillator systems, nature 
is also full with several amazing examples in this sense \cite{glass,winfree,neda1}. 
Synchronization in all these systems appears as a result of some specific coupling between the units. This coupling can be local or global, and can be realized through a phase-difference minimizing force \cite{kuramoto,gomez} or through the pulses emitted and detected by the oscillators \cite{bottani,pikovsky}.  In most of these synchronizing systems there is a clear driving force favoring synchronization, and in such way the appearance of this collective behavior is somehow trivial. In the present work however, a nontrivial synchronization will 
be presented. This weak synchronization (phase-locking) appears as a co-product of a 
simple collective optimization rule. 

One well-known
phenomena which inspired us in this work is the collective behavior and synchronization 
of fireflies \cite{fireflies}.  Although our aim here is not to model fireflies,
the oscillators ("electronic fireflies") considered in our system are somehow similar to them:
they are capable of emitting light-pulses and detecting the light-pulse of the others.
In this sense our system is similar to an ensemble of fireflies although the coupling between 
the units is different. From another perspective, the oscillators behave like pulse-coupled 
"integrate and fire" type neurons \cite{bottani,pikovsky,neda2}.  Contrary to the classical integrate and fire oscillators, in the considered system an inhibitory type global interaction is considered. This means that the firing 
of one oscillator delays (and not advances) the phase of all the others. This system does not necessarily 
favor synchronization, it is rather designed to keep a desired $W$ light intensity in the system. 
This light intensity is controlled by a firing threshold parameter $G$ imposed globally on the oscillators. 
Surprisingly, as a co-product of this simple rule, 
for certain region of the firing
threshold parameter phase-locking and complex patterns in the flashing sequence
of the oscillators will appear. We believe that such dynamical laws could be realistic for many 
biological systems.

The studied system will be described in more details in the following section. The used electronic device will
be briefly presented and the obtained non-trivial collective behavior will be studied. 
In order to get more confidence in the observed non-trivial results computer simulations were
also performed.

\section{The experimental setup}

\begin{figure}
\includegraphics[width=0.45\textwidth]{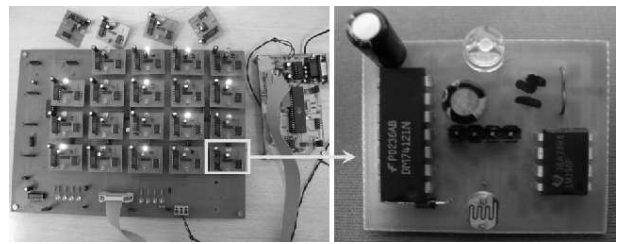}
\caption{Experimental setup. The photo on the left shows the "electronic fireflies" (oscillators) 
placed on the circuit board. The photo on the right shows one oscillator.}
\label{setup}
\end{figure}

\begin{figure}
\includegraphics[width=0.45\textwidth]{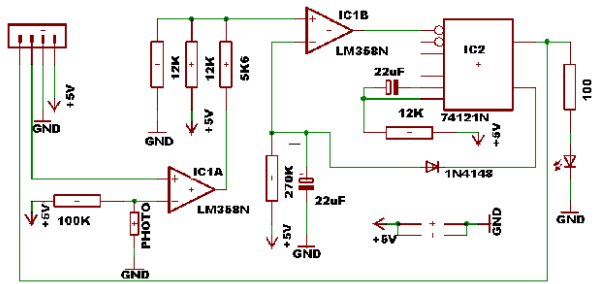}
\caption{Circuit diagram of one oscillator.}
\label{circuit}
\end{figure}

The constructed units are integrate and fire type oscillators \cite{bottani}
with a modified interaction rule.  Their coupling and communication is through light, 
the units are capable of emitting and detecting light-pulses.
The oscillators are practically realized by a relatively simple circuit, 
the main active elements being a photoresistor and a Light Emitting Diode (LED). 
Each oscillator, $i$,  has a characteristic voltage $U_i$, which depends on the resistance,  
$R_i$, of its photoresistor. The global light intensity influences the value of $R_i$ 
in the following sense: when the light intensity increases $R_i$ decreases, leading to a decrease in $U_i$.
In the system there is a global controllable parameter $G$,
identical for all oscillators. By changing
the parameter $G$, one can control the average light intensity output, $W$, of the whole system.
If the voltage of the oscillator grows above this threshold ($U_i>G$) the oscillator will fire, 
this meaning its LED will flash. This flash occurs only if a minimal time period $T_{min_i}$ has 
expired since the last firing. The oscillator has also a maximal period, meaning if no flash occurred 
in time $T_{max_i}$, then the oscillator will surely fire. In 
laymen terms firing is favored by darkness and the value of the controllable $G$ parameter characterizes 
the  ''darkness level'' at which firing should occur. Through this simple rule the $G$ parameter
controls the average light intensity output of the system.
The technical realization of the above dynamics is illustrated in Fig. \ref{circuit}. 
After the system has fired the $22\ \mu$F capacitor is completely discharged by the negative pulse from 
the inverted output of the monostable. As soon as the light flash ended, the same capacitor 
will start charging from the current flow through the $270$ K$\Omega$ resistor. The IC1B comparator 
will trigger another flash as soon as the potential on the mentioned capacitor will overcome 
the value fixed by the group of three resistors on its positive input (the firing threshold). 
Two of the resistors are connected to constant potentials (ground and +5 V), the third resistor is 
connected to the output of the second comparator IC1A which will have a value depending on the 
ratio between the reference voltage and a certain amount of light measured by the photo resistor.
The flash time is determined by the second capacitor together with the $12$ K$\Omega$ 
resistor connected to the monostable. The photoresistor has a relatively low reaction time around $40$ ms, 
while the minimal and maximal period of firing are around $800$ ms and $2700$ ms. 
The time of one flash is around $200$ ms. 

The oscillators are placed on a circuit board in the form of a square lattice (see Figure \ref{setup}). 
The maximal number of oscillators which can be included are $24$. A computer interface and program
controls the $G$ threshold parameter and allows us to get information automatically about the states 
of all oscillators. The state of an oscillator is recorded as $0$ if the oscillator does not emit light and
$1$ when the oscillator fires (emits light). Whenever the state of the
oscillator system changes, the program writes in a file the corresponding time with a precision of milliseconds 
and the new states of the units. 

To obtain an enhanced global interaction the whole system is placed inside a closed box. The box has mat 
glass mirror walls to uniformly disperse the light-pulses in the box. A graphical interface allows to 
visually control the state of the units.

In order to fully  understand the behavior of the system one has to accept that the coupling 
between pairs of oscillators are not exactly of the same strength.
Also, the characteristic electronic parameters differ slightly ($2-10\%$) among the units.

\section{Collective behavior}

At constant light intensity one unit behaves as a simple stochastic oscillator. 
Whenever the $G$ threshold is under a given $G_c$ value the oscillator
will fire with its minimal period and above $G_c$ with its maximal period. 
$G_c$ depends of course on the imposed light intensity. 
Considering more oscillators ($i=1,\dots ,n$) and by letting them interact,  
interesting collective behavior appears for a certain range of the $G$ threshold parameter.

Due to the inhibitory nature of the considered interaction, during the firing of oscillator $i$
the characteristic voltages of the others ($U_j, j\neq i$) will decrease.  If the $G$ parameter 
is so small, that under this condition the other oscillators 
can still fire ($U_j>G$), than all oscillators will fire in an uncorrelated manner. Each of them will be firing at 
its own $T_{min_j}$ period and the interaction is thus not efficient. In such case no collective behavior 
can be observed. 

Increasing the value of $G$ will make the pulse-like interaction efficient. 
The oscillators will avoid firing simultaneously and a simple
phase-locking phenomenon appears. The pulse of one unit (let us assume $i$) delays the firing of the 
others by decreasing their voltages below the threshold:  $U_j<G, j\neq i$. Due to the tiny 
differences 
in the coupling between the pairs (caused for example by different distances) and in the parameters 
of the electronic elements, the $U_j$ voltages are different. The immediate consequence of this is 
that the next firing will occur most probably in the oscillator with the highest voltage 
(counting of course only those oscillators, which are already capable of firing). This oscillator is the 
one which was influenced the less by the light-pulse of the previous firing. 
If the total combined time of firing for the $n$ oscillators is smaller than the period 
$T_{max}$ the result is that after very short time phase-locking appears and a firing chain 
(with period $T\in [T_{min},T_{max}]$) will form, each oscillator firing in a well-defined sequence. 
For a given  system and a fixed $G$ threshold this stable pattern is always the same. 
If the total time of firing of the $n$ oscillators exceeds $T_{max}$,  
the firing pattern will be much longer and more complex. 

Increasing further and over a limit the $G$ threshold parameter the previously discussed weak synchronization 
(phase-locking) disappears. In this case the voltages of 
all oscillators are much smaller than the threshold value $U_i<G$, so the firing of a unit can not influence 
the others. All oscillators will fire with their own $T_{max_i}$ period and no 
interesting collective behavior is observed. Again, the interaction is not efficient.

The collective behavior of the system can be easily analyzed by plotting a kind of {\em 
phase-histogram} for the oscillator ensemble. Choosing a reference oscillator, the relative phases of all
the others are defined by measuring the time difference between their pulse
and the last pulse emitted by the reference oscillator. Studying these time-delays during a 
longer time period a histogram is constructed for their distribution. This histogram shows how frequently a 
given time-delay occurred and gives thus a hint whether a constant firing pattern is formed or not.
 
\begin{figure}
\includegraphics[width=0.45\textwidth]{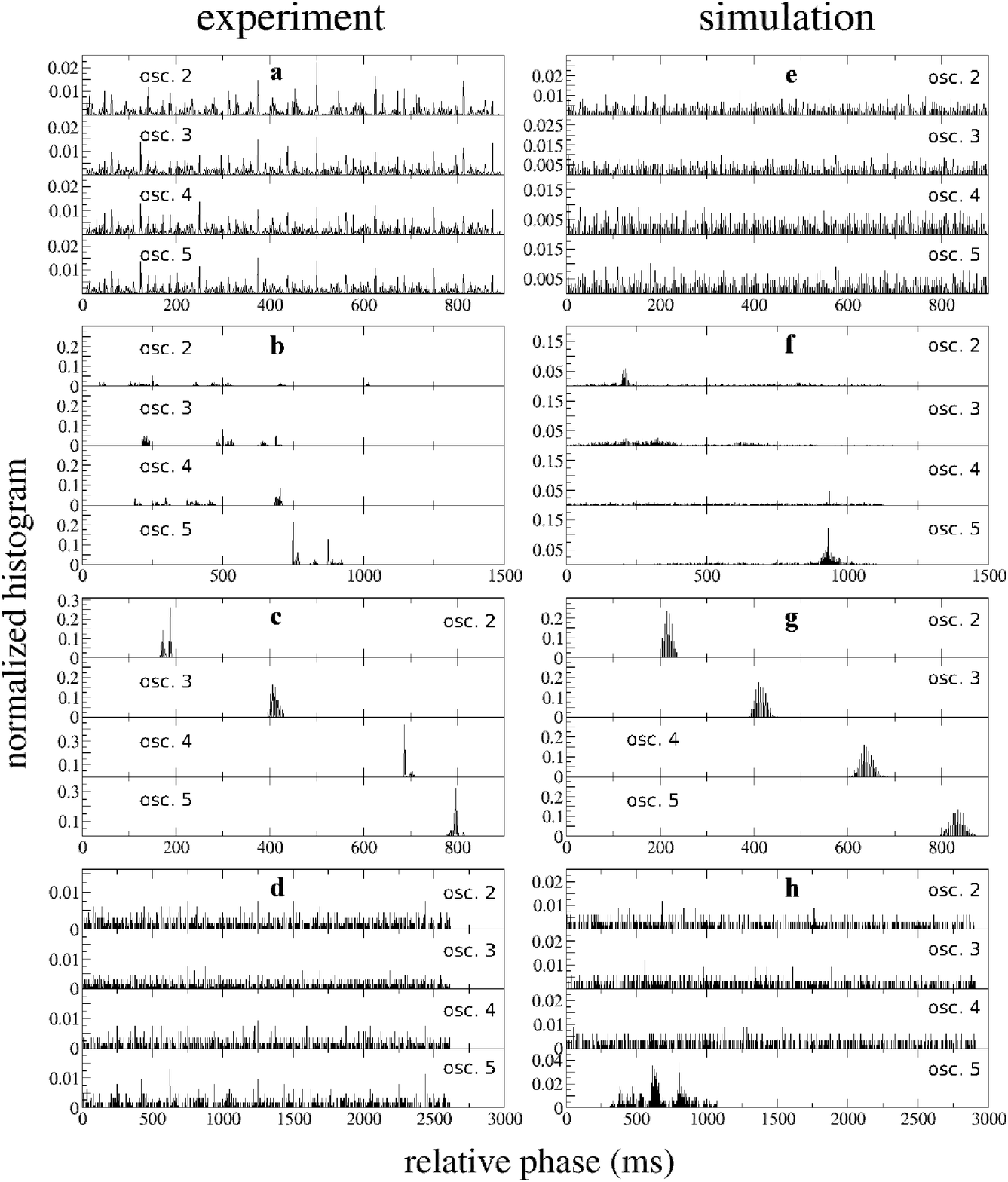} 
\caption{Relative phase histogram for $n=5$ oscillators. Experimental results are in the 
first column, and the corresponding simulation results are in the second column. Figures a) and e) 
are for $G=500$ mV; figure  b) and f) are for $G=2000$ mV; figures c) and g) are for $G=3000$ mV, and figures d) and h)  are for $G=4200$ mV.}
\label{5led}
\end{figure}

Experimental and computer simulated results for the phase-histogram confirm the above presented
scenario of the collective behavior. As an example, on Fig. \ref{5led}, 
results obtained on a relatively small system with $n=5$ oscillators are shown.   
In the first column of Fig. \ref{5led} (figures a, b, c and d), experimental results for four different values 
of the $G$ threshold are plotted. For a small threshold parameter ($G=500$ mV), 
no self-organization appears (Fig. \ref{5led}a). Due to the fact that the characteristic time-periods 
of the oscillators are slightly different, almost all values will occur with 
the same probability in the phase-histogram. Beginning with $G=1300$ mV
a kind of order begins to 
emerge, and a trend towards the self-organization of the oscillator pulses is observed (e.g. Fig. \ref
{5led}b for $G=2000$ mV).  
In the neighborhood of $G=3000$ mV threshold value (Fig. \ref{5led}c) clear phase-locking appears. 
One can observe that a stable 
firing pattern has formed,  each oscillator has an almost exact phase relative to 
the reference oscillator. For an even higher value (e.g. $G=4200$ mV), disorder sets in
again, phase-locking disappears and all oscillators fire independently with their own maximal period 
(Fig. \ref{5led}d).

In the second column of  Fig. \ref{5led} we present the corresponding simulation results. 
In simulations the parameters of the oscillators are defined as following: 
the average minimal time period is $T_{min_i}=900$ ms, the average 
maximal period $T_{max_i}=2700$ ms,  and the average flashing time $T_{flash} = 200$ ms. 
For an easier comparison, the values are chosen to be similar with the real experimental data. 
We considered a uniform distribution of the oscillators parameter around these average values using a $\pm 50$ ms
interval for $T_{min}$ and $T_{max}$ and a $\pm 20$ ms interval for $T_{flash}$.  
One could argue of course that a Gaussian distribution would be much more appropriate, but given the fact 
that we simulate here relatively small number of oscillators the exact statistics is irrelevant. 
Considering some deviations from the average is however 
important in order to reproduce the collective behavior of the system. 
An uncorrelated noise in time is also considered. This will randomly shift the 
$T_{min}$, $T_{max}$ and $T_{flash}$ periods of each oscillator at each cycle. Again, 
a uniform distribution on a $\pm 20$ ms interval was considered. 
The characteristic voltages of the oscillators are set to be in the interval $4100\pm 100$ mV in dark, 
$2100 \pm 100$ mV when one single LED is flashing and $1050 \pm 100$ mV when two LEDs are flashing simultaneously. 
Whenever $k$ LEDs are simultaneously flashing the characteristic voltages of the others are considered to be
$2100/k \pm 100$ mV, however for $n=5$ oscillators only very rarely happens to have more than two oscillators 
simultaneously firing. This values were chosen to 
approximately match the experimental ones, and we do not try here to give a theoretical model for 
the nonlinear behavior of the photoresistor. Fluctuations in time and among the parameters of the oscillators 
are again included. Differences in the strength of the coupling between pairs of oscillators are however neglected. 
Using these parameters, it is assumed that each oscillator can flash whenever its voltage exceeds the 
threshold $G$. The flashing cannot occur earlier than $T_{min_i}$ or later than $T_{max_i}$ 
relatively to its last firing. On Fig. \ref{5led}e.,f.,g. and h., 
the simulated phase-histograms of the 
oscillators are plotted and compared with the corresponding experimental data. The observed experimental results, 
including the non-trivial synchronization (phase-locking),  were successfully reproduced.  

\begin{figure}
\includegraphics[width=0.45\textwidth]{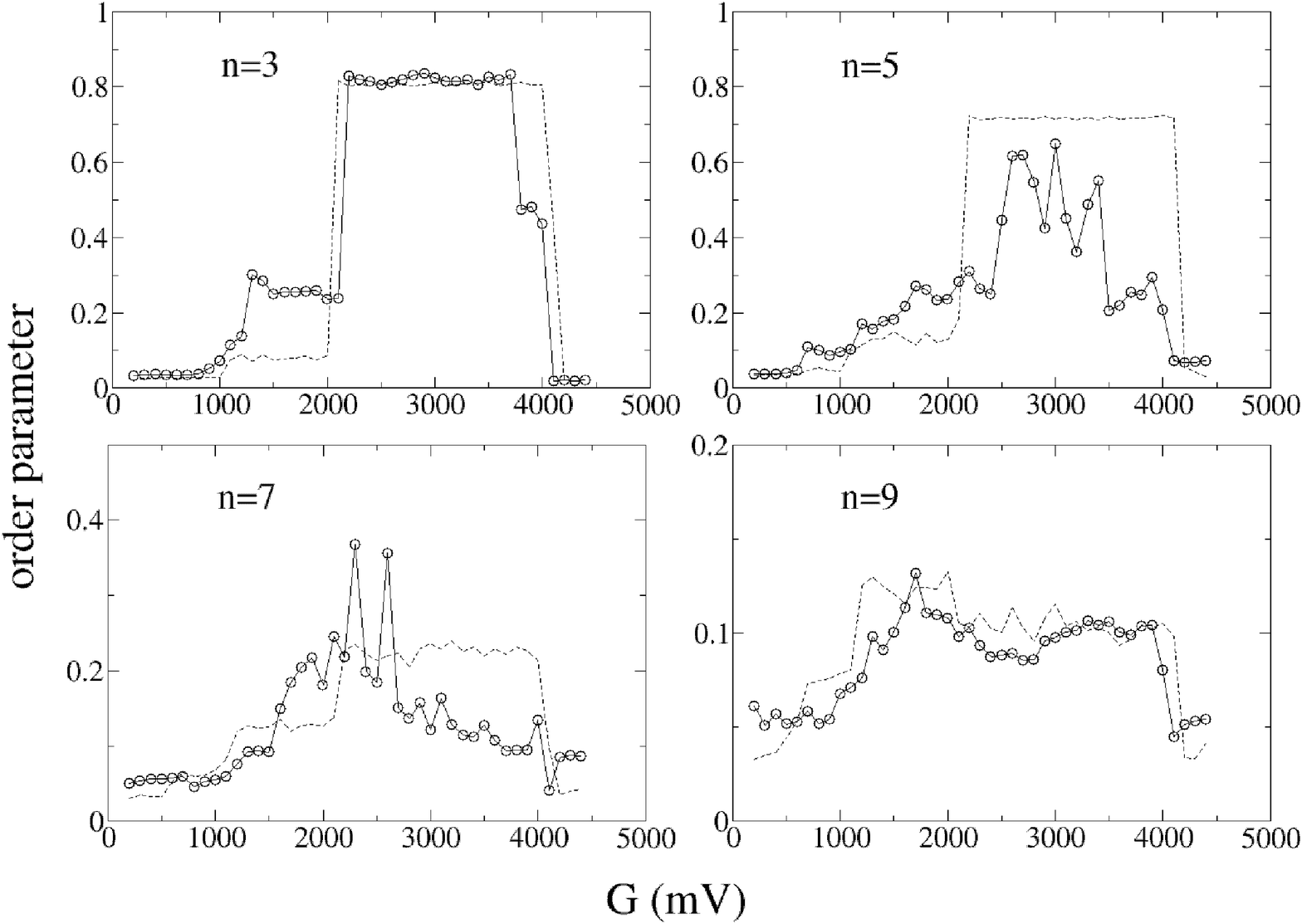}
\caption{Order parameters calculated from experimental (circles) and simulation (dashed line) results 
plotted as a function of the $G$ threshold. Systems with $n=3,5,7,9$ oscillators are considered.}
\label{orderparameter}
\end{figure}

It is also possible to define a kind of order-parameter that characterizes the observed 
synchronization 
level. Our method for calculating this is the following:\\
1) A reference oscillator $k$ is chosen and the phases of all oscillators are calculated relative to 
this oscillator.
2) Let  $h_i (f)$ denote the value of the normalized phase-histogram for oscillator $i$ 
($i={1,\dots,n}, i\neq k$) corresponding to phase difference (time difference) value $f$. Since we
have a normalized histogram, $h_i(f)\in [0,1]$ gives the occurrence probability of 
phase difference value $f$ during the measurement ($\sum_f{h_i(f)}=1$).\\
3) A window of width $a$ is defined (we have chosen $a=30$ ms). Shifting the window with 
$\Delta f=1$ ms step, for each discretized value of $f$ the sum $H_i(f)=\sum_{j=f-a/2}^{f+a/2} h_i(j)$ 
is calculated for each oscillator $i$.\\
4) Let $r_k$ denote the difference between the maximum and minimum value of $H_i(f)$ averaged 
over all oscillators:  $r_k=\frac{1}{n-1} \sum_{i=1,(i\ne k)}^{n} max(H_i)-min(H_i)$.\\
5) Items 1-4 are repeated considering each oscillator in the system as reference oscillator. \\ 
Finally, an averaging is performed over all the obtained $r_k$ values ($k={1,\dots,n}$). 
The final order parameter is calculated thus as $r=\langle r_k \rangle_k$. Averaging as a 
function of the reference oscillator  
is beneficial in order to get a smoother curve 
when only partial phase locking is detected (Figure
\ref{5led}b.). In such cases the phase-diagrams 
 are very sensible on the choice of the reference oscillator.  

On Fig. \ref{orderparameter} the $r$ order parameter is plotted as a function of the 
$G$ threshold value.  Systems with $n=3,5,7$ and $9$ oscillators are considered. 
Experimental (circles) and simulation results (dashed line) are again in good agreement.
The figure also illustrates that for an intermediate $G$ interval value phase-locking appears. 
This weak synchronization is better ($r$ is bigger) when there are less units in the system. 
One obvious reason for this is that by increasing $n$ the total time of firing of the oscillators 
will increase and slowly exceed the value $T_{max}$. As a result of this the firing pattern will 
change from a simple "firing chain" to a much longer and more complicated pattern, decreasing the 
value of the order parameter. 

From Fig. \ref{orderparameter} it is also observable that 
the experimental results show more intensive fluctuations. The reason for this is probably 
the complex noise present in the system.

\section{Conclusion}

A system of electronic oscillators communicating through light-pulses was studied. The units were designed to
optimize the average light intensity of the emitted light-pulses, and no direct driving force
favoring synchronization was considered. Although our experiments focused on relatively small systems 
(up to $24$ oscillators) interesting and rich collective behavior was observed. As a nontrivial result it was 
found that the inhibitory coupling  induced a partial phase-locking for a certain interval of the 
controllable threshold parameter. This weak synchronization was realized by complex flashing patterns 
of the units. We believe that this study inspires further interesting research projects in which 
separately programmable oscillators will be studied with various interaction rules. Many other 
interesting collective behaviors can be obtained by controlling individually 
the parameters of the units, their interaction rule and the type of coupling between them. 
Such systems could also yield a new approach to 
unconventional computing, being in many sense similar with the presently developed CNN computers \cite{marek}.

{\bf Acknowledgments.}
Work supported from a Romanian CNCSIS No.1571 research  grant (contract 84/2007) 
and a Hungarian ONR grant (N00014-07-1-0350).

\end{document}